\documentclass[a4paper]{llncs}

\usepackage{graphicx}
\usepackage{amsmath}
\usepackage{amssymb}
\usepackage[pagebackref=false,breaklinks=true,colorlinks,bookmarks=false,urlcolor=blue,linkcolor=blue,citecolor=blue]{hyperref}

\usepackage{mathtools}
\usepackage{xspace}
\usepackage{multirow}
\usepackage{xfrac}
\usepackage{booktabs}  

\usepackage{fancyvrb}  
\usepackage{adjustbox}

\usepackage{times}        

\usepackage[scaled=0.9]{inconsolata}  

\newcommand{\keywords}[1]{\par\addvspace\baselineskip
\noindent\keywordname\enspace\ignorespaces#1}

\begin{document}

\mainmatter

\title{A User-Friendly Hybrid Sparse Matrix Class in C++}
\titlerunning{A User-Friendly Hybrid Sparse Matrix Class in C++}

\author{Conrad Sanderson~\inst{1,3,4} \and  Ryan Curtin~\inst{2,4}}
\authorrunning{Sanderson, Curtin}

\institute
  {
  Data61, CSIRO, Australia\\
  \and
  Symantec Corporation, USA\\
  \and
  University of Queensland, Australia\\
  \and
  Arroyo Consortium
  ~\textcolor{white}{\thanks{\scriptsize {\bf Published in:} Lecture Notes in Computer Science (LNCS), Vol.~10931, pp.~422-430, 2018.\protect\\ \href{https://doi.org/10.1007/978-3-319-96418-8_50}{\tt https://doi.org/10.1007/978-3-319-96418-8\_50}}}
  }

\maketitle


\vspace{-3ex}
\begin{abstract}

When implementing functionality which requires sparse matrices,
there are numerous storage formats to choose from,
each with advantages and disadvantages.
To achieve good performance, several formats may need to be used in one program,
requiring explicit selection and conversion between the formats.
This can be both tedious and error-prone, especially for non-expert users.
Motivated by this issue, we present a user-friendly sparse matrix class for the \mbox{C++} language,
with a high-level application programming interface deliberately similar to the widely used \mbox{MATLAB} language.
The class internally uses two main approaches to achieve efficient execution:
(i) a hybrid storage framework, which automatically and seamlessly switches between three underlying storage formats
(compressed sparse column, coordinate list, Red-Black tree)
depending on which format is best suited for specific operations,
and
(ii) template-based meta-programming to automatically detect and optimise execution of common expression patterns.
To facilitate relatively quick conversion of research code into production environments,
the class and its associated functions provide a suite of essential sparse linear algebra functionality
(eg., arithmetic operations, submatrix manipulation)
as well as high-level functions for sparse eigendecompositions and linear equation solvers.
The latter are achieved by providing easy-to-use abstractions of the low-level ARPACK and \mbox{SuperLU} libraries.
The source code is open and provided under the permissive Apache~2.0 license,
allowing unencumbered use in commercial products.

\vspace{-1.5ex}
\keywords{numerical linear algebra, sparse matrix, C++ language}

\end{abstract}

\vspace{-6ex}
\section{Introduction}
\vspace{-1.2ex}

Modern scientific computing often requires working with data so large it cannot fully fit in working memory.
In many cases, the data can be represented as sparse, allowing users to work with matrices of extreme size with few nonzero elements.
However, converting code from using dense matrices to using sparse matrices is not always straightforward.

Existing open-source frameworks may provide several separate sparse matrix
classes, each with their own data storage format.  For instance,
SciPy~\cite{SciPy} has 7 sparse matrix classes: {\tt bsr\_matrix}, {\tt
coo\_matrix}, {\tt csc\_matrix}, {\tt csr\_matrix}, {\tt dia\_matrix}, {\tt
dok\_matrix}, and {\tt lil\_matrix}.
Each storage format is best suited for efficient execution of a specific set of operations (eg.,~matrix multiplication vs.~incremental matrix construction).
Other frameworks may provide only one sparse matrix class, with severe runtime penalties if it is not used in the right way.
This can be challenging and bewildering for users who simply want to create
and use sparse matrices, and do not have the expertise (or desire) to
understand the advantages and disadvantages of each format.
To achieve good performance, several formats may need to be used in one program,
requiring explicit selection and conversion between the formats.
This plurality of sparse matrix classes complicates the programming task,
increases the likelihood of bugs, and adds to the maintenance burden.

Motivated by the above issues, we present a user-friendly sparse matrix class for the C++ language,
with a high-level application programming interface (function syntax) that is deliberately similar to MATLAB.
The sparse matrix class uses a hybrid storage framework, which \mbox{\it automatically} and \mbox{\it seamlessly} switches between three data storage formats,
depending on which format is best suited for specific operations:
{\bf (i)}~Compressed Sparse Column (CSC), used for efficient fundamental arithmetic operations such as matrix multiplication and addition, as well as efficient reading of individual elements;
{\bf (ii)}~Co-Ordinate List (COO), used for facilitating operations involving bulk coordinate transformations;
{\bf (iii)}~Red-Black Tree (RBT), used for both robust and efficient incremental construction of sparse matrices
(ie.,~construction via setting individual elements one-by-one, not~necessarily in order).
To further promote efficient execution, the class exploits C++ features such as template meta-programming
to provide a compile-time expression evaluator,
which can automatically detect and optimise common mathematical expression patterns.

The sparse matrix class provides an intuitive interface that is very close to
a typical dense matrix API;
this can help with rapid transition of dense-specific code to sparse-specific code.
In addition, we demonstrate that the overhead of the hybrid format is minimal,
and that the format is able to choose the optimal representation for a variety of sparse linear algebra tasks.
This makes the format and implementation suitable for real-world prototyping and production usage.

Although there are many other sparse matrix implementations in existence, to our
knowledge ours is the first to offer a unified interface with automatic format
switching under the hood.  Most toolkits are limited to either a single format
or multiple formats the user must manually convert between.  As mentioned
earlier, SciPy contains no fewer than seven formats, and the comprehensive
SPARSKIT package~\cite{saad1990sparskit} contains 16.  In these toolkits the
user must manually convert between formats.  On the other hand, both MATLAB and
GNU Octave~\cite{octave} contain sparse matrix implementations, but they supply
only the CSC format, meaning that users must write their code in special
ways to ensure its efficiency~\cite{matlab_fast_matrix}.

The source code for the sparse matrix class and its associated functions
is included in recent releases of the cross-platform and open-source Armadillo linear algebra library~\cite{Armadillo_JOSS_2016},
available from \url{http://arma.sourceforge.net}.
The code is provided under the permissive Apache~2.0 license~\cite{Rosen_2004},
allowing unencumbered use in commercial products.

We continue the paper as follows.
In Section~\ref{sec:functionality} we overview the functionality provided by the sparse matrix class and its associated functions.
In Section~\ref{sec:formats} we briefly describe the underlying storage formats used by the class, and the tasks that each of the formats is best suited for.
Section~\ref{sec:experiments} provides an empirical evaluation showing the advantages of the hybrid storage framework and automatic expression optimisation.
The salient points and avenues for further exploration are summarised in Section~\ref{sec:conclusion}.

\vspace*{-0.7em}
\section{Functionality}
\label{sec:functionality}
\vspace*{-0.5em}

To allow prototyping directly in C++ as well as to facilitate relatively quick conversion of research code into production environments,
the sparse matrix class and its associated functions provide a user-friendly suite of essential sparse linear algebra functionality,
including fundamental operations such as addition, matrix multiplication and submatrix manipulation.
Various sparse eigendecompositions and linear equation solvers are also provided.
C++ language features such as overloading of operators (eg.,~{\tt *}~and~{\tt +})~\cite{Stroustrup_2013}
are exploited to allow mathematical operations with matrices to be expressed in a concise and easy-to-read manner.
For instance, given sparse matrices \texttt{A}, \texttt{B}, and \texttt{C},
a~mathematical expression such as

\centerline{\texttt{D = $\frac{1}{2}$(A + B) $\cdot$ C\textsuperscript{T}}}

\noindent
can be written directly in C++ as

\vspace*{0.5em}
\centerline{\texttt{sp\_mat D = 0.5 * (A + B) * C.t()};}
\vspace*{0.5em}

Low-level details such as memory management are hidden, allowing the user to concentrate effort on mathematical details.
Table~\ref{tab:function_list} lists a subset of the available functionality for the sparse matrix class, {\tt sp\_mat}.

The sparse matrix class uses a delayed evaluation approach,
allowing several operations to be combined to reduce the amount of computation and/or temporary objects.
In contrast to brute-force evaluations, delayed evaluation can provide considerable performance improvements as well as reduced memory usage.
The delayed evaluation machinery is accomplished through template meta-programming~\cite{Vandevoorde_2018},
where a type-based signature of a set of consecutive mathematical operations is automatically constructed.
The C++ compiler is then induced to detect common expression subpatterns at compile time, and selects the corresponding optimised implementations.
For example, in the expression \mbox{\footnotesize\tt trace(A.t() * B)}, the explicit transpose and time-consuming matrix multiplication are omitted;
only the diagonal elements of \mbox{\footnotesize\tt A.t() * B} are accumulated.

Sparse eigendecompositions and linear equation solutions are accomplished
through integration with low-level routines in the de facto standard ARPACK~\cite{lehoucq1998arpack} and SuperLU libraries~\cite{li2005overview}.
The resultant high-level functions automatically take care of the cumbersome and error-prone low-level management required with these libraries.

\begin{table}[!tb]
\footnotesize
\centering
\begin{tabular}{ll}
\toprule
{\bf Function} & {\bf Description} \\
\hline
\texttt{~~~sp\_mat X(100,200)}        & Declare sparse matrix with 100 rows and 200 columns \\
\texttt{sp\_cx\_mat X(100,200)}       & As above, but use complex elements \\
\texttt{X(1,2) = 3}                   & Assign value 3 to element at location (1,2) of matrix {\it X} \\
\texttt{X = 4.56 * A}                 & Multiply matrix {\it A} by scalar \\
\texttt{X = A + B}                    & Add matrices {\it A} and {\it B} \\
\texttt{X = A * B}                    & Multiply matrices {\it A} and {\it B} \\
\texttt{X = kron(A, B)}               & Kronecker tensor product of matrices {\it A} and {\it B} \\
\texttt{X( span(1,2), span(3,4) )}    & Provide read/write access to submatrix of {\it X} \\
\texttt{X.diag(k)}                    & Provide read/write access to diagonal {\it k} of {\it X} \\
\texttt{X.print()}                    & Print matrix {\it X} to terminal \\
\texttt{X.save(filename, format)}     & Store matrix {\it X} as a file \\
\texttt{speye(rows, cols)}            & Generate sparse matrix with values on diagonal set to one \\
\texttt{sprandu(rows, cols, density)}
                                      & Generate sparse matrix with random non-zero elements \\
\texttt{sum(X, dim)}                  & Sum of elements in each column~({\it dim=0}) or row ({\it dim=1}) \\
\texttt{min(X, dim); max(X, dim)}     & Obtain extremum value in each col.~({\it dim=0}) or row ({\it dim=1}) \\
\texttt{X.t()}~~or~~\texttt{trans(X)} & Return transpose of matrix {\it X} \\
\texttt{repmat(X, rows, cols)}        & Replicate matrix {\it X} in block-like fashion \\
\texttt{norm(X, p)}                   & Compute {\it p}-norm of vector or matrix {\it X} \\
\texttt{normalise(X, p, dim)}         & Normalise each col.~({\it dim=0}) or row ({\it dim=1}) to unit {\it p}-norm \\
\texttt{trace(A.t() * B)}             & \scalebox{0.95}{Compute trace {\bf omitting} explicit transpose and multiplication} \\
\texttt{eigs\_gen(eigval, eigvec, X, k)}
                                      & Compute {\it k} largest eigenvalues and eigenvectors of matrix {\it X} \\
\texttt{svds(U, s, V, X, k)}          & Compute {\it k} singular values and singular vectors of matrix {\it X} \\
\texttt{X = spsolve(A, b)}            & Solve sparse system {\it Ax = b} for {\it x} \\
\bottomrule
\end{tabular}
\vspace{0.5ex}
\caption
  {
  Selected functionality of the sparse matrix class, with brief descriptions.
  See {\href{http://arma.sourceforge.net/docs.html\#SpMat}{\mbox{\tt http://arma.sourceforge.net/docs.html\#SpMat}}} for more detailed documentation.
  Several optional additional arguments have been omitted for brevity.
  }
\label{tab:function_list}
\vspace{-4ex}
\end{table}

\section{Underlying Sparse Storage Formats}
\label{sec:formats}

The three underlying storage formats (CSC, COO, RBT) were chosen
so that the sparse matrix class can achieve overall efficient execution of the following five main use cases:
(i)~incremental construction of sparse matrices via quasi-ordered insertion of elements,
where each new element is inserted at a location that is past all the previous elements according to column-major ordering;
(ii)~flexible ad-hoc construction or element-wise modification of sparse matrices via unordered insertion of elements,
where each new element is inserted at a random location;
(iii)~operations involving bulk coordinate transformations;
(iv)~multiplication of dense vectors with sparse matrices;
(v)~multiplication of two sparse matrices.

Below we briefly describe each storage format and its limitations.
We use $N$ to indicate the number of non-zero elements of the matrix,
while \texttt{n\_rows} and \texttt{n\_cols} indicate the number of rows and columns, respectively.

\subsection{Compressed Sparse Column}

In the CSC format~\cite{saad1990sparskit}, three arrays are used:
(i)~the {\it values} array, which is a contiguous array of $N$ floating point numbers holding the non-zero elements,
(ii)~the {\it row indices} array, which is a contiguous array of $N$ integers holding the corresponding row indices (ie., the $n$-th entry contains the row of the $n$-th element),
and
(iii)~the {\it column offsets} array, which is a contiguous array of $\texttt{n\_cols}+1$ integers holding offsets to the {\it values array},
with each offset indicating the start of elements belonging to each column.
Let us denote the $i$-th entry in the column offsets array as~$c[i]$, the $j$-th entry in the row indices array as~$r[j]$, and the $n$-th entry in the values array as~$v[n]$.
All arrays use zero-based indexing, ie.,~the initial position in each array is denoted by~$0$.
Then, $v[\hspace{0.25ex}c[i]\hspace{0.25ex}]$ is the first element in column $i$, and $r[\hspace{0.25ex}c[i]\hspace{0.25ex}]$ is the corresponding row of the element.
The number of elements in column $i$ is determined using $c[i\!+\!1] - c[i]$, where, by definition, $c[0]$ is always $0$ and $c[\texttt{n\_cols}]$ is equal to $N$.

The CSC format is well-suited for sparse linear algebra operations such as summation and vector-matrix multiplication.
It is also suited for operations that do not change the structure of the matrix, such as element-wise operations on the nonzero elements.
The format also affords relatively efficient random element access;
to locate an element (or determine that it is not stored),
a single lookup to the beginning of the desired column can be performed,
followed by a binary search to find the element.

The main disadvantage of CSC is the effort required to insert a new element.
In the worst-case scenario, memory for three new larger-sized arrays (containing the values and locations) must first be allocated,
the position of the new element determined within the arrays,
data from the old arrays copied to the new arrays,
data for the new element placed in the new arrays,
and finally the memory used by the old arrays deallocated. 
As the number of elements in the matrix grows, the entire process becomes slower.

There are opportunities for some optimisation, such as using oversized storage to reduce memory allocations,
where a new element past all the previous elements can be readily inserted.
It is also possible to perform batch insertions with some speedup
by first sorting all the elements to be inserted and then merging with the existing data arrays.
While the above approaches can be effective, they require the user to explicitly deal with low-level storage details
instead of focusing on high-level functionality.

The CSC format was chosen over the related Compressed Sparse Row (CSR) format~\cite{saad1990sparskit}
for two main reasons:
(i) to ensure compatibility with external libraries such as the SuperLU solver~\cite{li2005overview},
and
(ii) to ensure consistency with the surrounding infrastructure provided by the Armadillo library,
which uses column-major dense matrix representation for compatibility with \mbox{LAPACK}~\cite{anderson1999lapack}.

\subsection{Coordinate List Representation }

The Coordinate List (COO) is a general concept where a list $L$ = $\left( l_1, l_2, \cdots, l_N \right)$ of \mbox{3-tuples} represents the non-zero elements in a matrix.
Each 3-tuple contains the location indices and value of the element, ie., $l$ = $\left( \texttt{row}, \texttt{column}, \texttt{value} \right)$.
The format does not prescribe any ordering of the elements,
and a linked list~\cite{Cormen_2009} can be used to represent $L$.
However, in a computational implementation geared towards linear algebra operations~\cite{saad1990sparskit},
$L$ is often represented as a set of three arrays:
(i)~the {\it values} array, which is a contiguous array of $N$ floating point numbers holding the non-zero elements of the matrix,
and the (ii)~{\it rows} and (iii)~{\it columns} arrays, which are contiguous arrays of $N$ integers, holding the row and column indices of the corresponding values.

The array-based representation of COO is related to CSC, with the main difference that for each element the column indices are explicitly stored.
As such, the COO format contains redundancy and is hence less efficient than CSC for representing sparse matrices.
However, in the COO format the coordinates of all elements can be directly read and modified in a batch manner,
which facilitates specialised/niche operations that involve bulk transformation of matrix coordinates (eg.,~circular shifts).
In the CSC format such operations are more time-consuming and/or more difficult to implement,
as the compressed structure must be taken into account.
The general disadvantages of the array-based representation of COO are similar as for the CSC format,
in that element insertion is typically a slow process.

\subsection{Red-Black Tree}

To address the problems with element insertion at arbitrary locations,
we first represent each element as a 2-tuple,
$l$ = $\left( \texttt{index}, \texttt{value} \right)$,
where $\texttt{index}$ encodes the location of the element as $\texttt{index}$ = $\texttt{row} + \texttt{column} \times \texttt{n\_rows}$.
This encoding implicitly assumes column-major ordering of the elements.
Secondly, rather than using a linked list or an array based representation,
the list of the tuples is stored as a Red-Black Tree (RBT),
a~self-balancing binary search tree~\cite{Cormen_2009}.

Briefly, an RBT is a collection of nodes, with each node containing the \mbox{2-tuple} described above and links to two children nodes.
There are two constraints:
(i)~each link points to a unique child node
and
(ii)~there are no links to the root node.
The ordering of the nodes and height of the tree
is explicitly controlled
so that searching for a specific index (ie., retrieving an element at a specific location)
has worst-case complexity of $\mathcal{O}(\log N)$.
Insertion and removal of nodes (ie.,~matrix elements), also has the worst-case complexity of $\mathcal{O}(\log N)$.
If a node to be inserted is known to have the largest index so far (eg.,~during incremental matrix construction),
the search for where to place the node can be omitted,
which in practice can considerably speed up the insertion process.

Traversing the tree in an ordered fashion (from the smallest to largest index)
is equivalent to reading the elements in column-major ordering.
This in turn allows the quick conversion of matrix data stored in RBT format into CSC format.
Each element's location is simply decoded via
\mbox{$\texttt{row}$ = $\texttt{index} \operatorname{~mod~} \texttt{n\_rows}$}
and
\mbox{$\texttt{column}$ = ${\lfloor}\texttt{index} / \texttt{n\_rows}{\rfloor}$},
with the operations accomplished via direct integer arithmetic on CPUs.

In our hybrid format, the RBT format is used for incremental construction of
sparse matrices, either in an ordered or unordered fashion,
and a subset of element-wise operations.
This in turn enables users to construct sparse matrices in the same way they
might construct dense matrices---for instance, a loop over elements to be
inserted without regard to storage format.

While the RBT format allows for fast element insertion,
it is less suited than CSC for efficient linear algebra operations.
The CSC format allows for exploitation of fast caches in modern CPUs
due to the consecutive storage of non-zero elements in memory~\cite{Mittal_2016}.
In contrast, accessing consecutive elements in the RBT format
requires traversing the tree (following links from node to node),
which in turn entails accessing node data that is not guaranteed to be consecutively stored in memory.
Furthermore, obtaining the column and row indices requires explicit decoding of the index stored in each node,
rather than a simple lookup in the CSC format.

\subsection{Automatically Switching Between Storage Formats}
\label{sec:auto}

To avoid the problems associated with selection and manual conversion between formats,
our sparse matrix class uses a hybrid storage framework that \mbox{\it automatically} and \mbox{\it seamlessly}
switches between the data storage formats described above.

By default, matrix elements are stored in CSC format.
When required, data in CSC format is internally converted to either the RBT or COO format, on which an operation or set of operations is performed.
The matrix is automatically converted (`synced') back to the CSC format the next
time an operation requiring the CSC format is performed.

The actual underlying storage details and conversion operations are completely hidden from the user,
who may not necessarily be knowledgeable about (or care to learn about) sparse matrix storage formats.
This allows for simplified code, which in turn increases readability and lowers maintenance.
In contrast, other toolkits without automatic format conversion can cause either slow execution (as a non-optimal storage format might be used),
or require many manual conversions.
As an example, Fig.~\ref{fig:manual_vs_automatic_conversion} shows a short Python program using the SciPy toolkit and a corresponding C++ program using the sparse matrix class.
Manually initiated format conversions are required for efficient execution in the
SciPy version; this causes both development time and code size to increase.

\begin{figure}[!tb]
\centering
\begin{adjustbox}{minipage=\columnwidth,scale={0.95}{0.95}}
\hrule
\hrule
\vspace{0.5ex}
\begin{minipage}{0.45\textwidth}
\begin{Verbatim}[fontsize=\footnotesize]
X = scipy.sparse.rand(1000, 1000, 0.01)

# manually convert to LIL format
# to allow insertion of elements
X = X.tolil()   
X[1,1]  = 1.23
X[3,4] += 4.56

# random dense vector
V = numpy.random.rand((1000))

# manually convert X to CSC format
# for efficient multiplication
X = X.tocsc()  
W = V * X
\end{Verbatim}
\end{minipage}
\hfill
\vline
\vline
\hfill
\begin{minipage}{0.45\textwidth}
\begin{Verbatim}[fontsize=\footnotesize]
sp_mat X = sprandu(1000, 1000, 0.01);

// automatic conversion to RBT format
// for fast insertion of elements

X(1,1)  = 1.23;
X(3,4) += 4.56;

// random dense vector
rowvec V(1000, fill::randu);

// automatic conversion of X to CSC
// prior to multiplication

rowvec W = V * X; 
\end{Verbatim}
\end{minipage}
\vspace{0.5ex}
\hrule
\hrule
\end{adjustbox}
\vspace{-1ex}
\caption
  {
  Left panel: a Python program using the SciPy toolkit, requiring explicit conversions between sparse format types to achieve efficient execution;
  if an unsuitable sparse format is used for a given operation, SciPy will emit {\it TypeError} or {\it SparseEfficiencyWarning}.
  Right panel: A corresponding C++ program using the sparse matrix class, with the format conversions automatically done by the class.
  }
\label{fig:manual_vs_automatic_conversion}
\end{figure}


\section{Empirical Evaluation}
\label{sec:experiments}

To empirically demonstrate the usefulness of the hybrid storage framework
and the template-based expression optimisation mechanism,
we have performed several experiments:
(i)~quasi-ordered element insertion, ie., incremental construction,
(ii)~unordered (random) insertion,
and
(iii)~calculation of {\small $\operatorname{trace}(A^T B)$},
where {\small $A$} and {\small $B$} are randomly generated sparse matrices.
In all cases the sparse matrices have a size of {\small 10,000$\times$10,000},
with four settings for the density of non-zero elements: {\small 0.01\%, 0.1\%, 1\%, 10\%}.

\begin{figure}[!tb]
\centering
\begin{minipage}{0.485\textwidth}
  \centering
  \includegraphics[width=1\textwidth,height=0.8\textwidth,keepaspectratio=false]{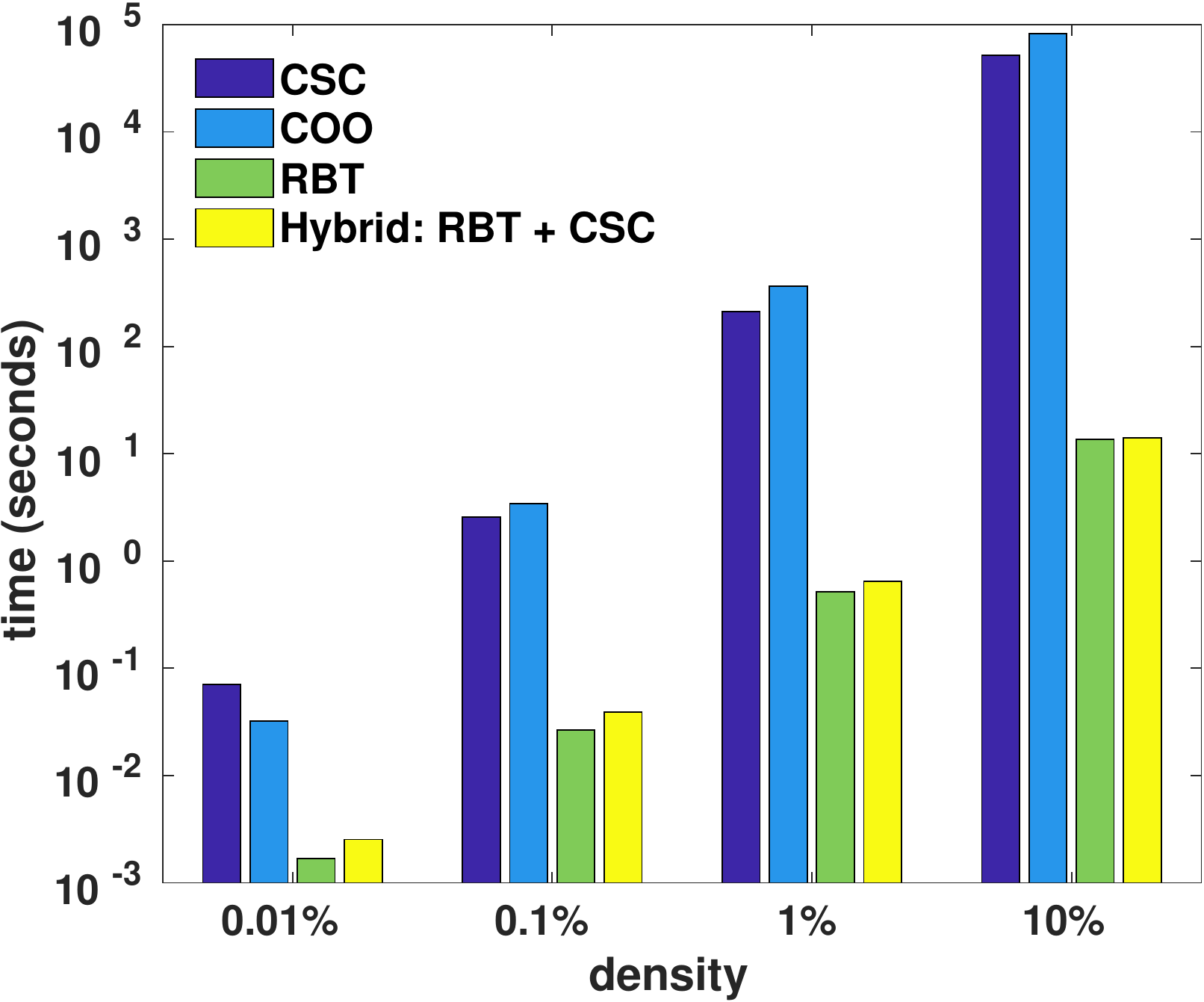}\\
  {\bf (a)}
  \end{minipage}%
  \hfill
  \begin{minipage}{0.485\textwidth}
  \centering
  \includegraphics[width=0.98\textwidth,height=0.8\textwidth,keepaspectratio=false]{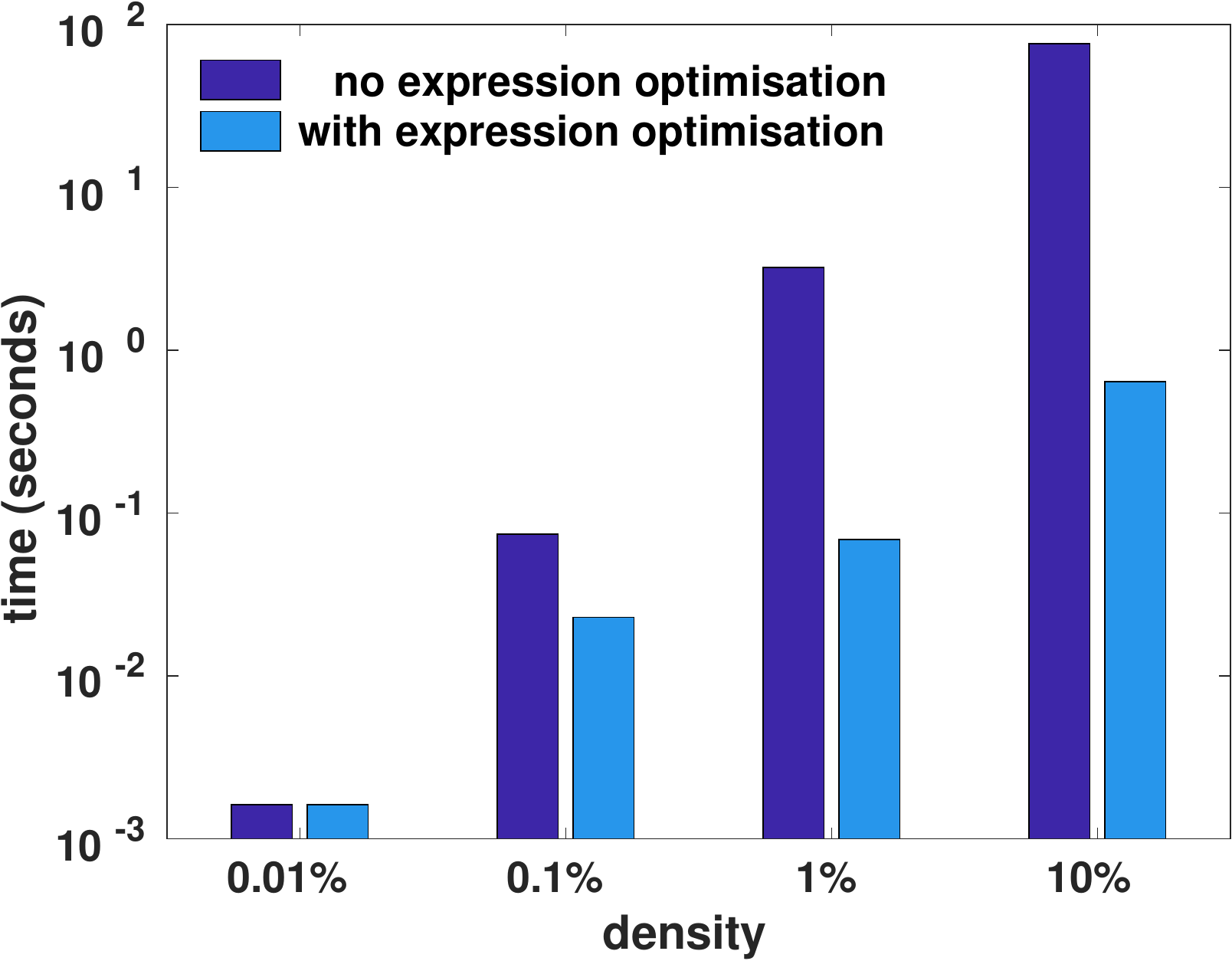}\\
  {\bf (b)}
  \end{minipage}
\caption
  {
  Time taken to
  {\bf (a)} insert elements at random locations into a sparse matrix to achieve various densities of non-zero elements,
  and
  {\bf (b)} calculate the expression {\tt trace(A.t()*B)},
  where {\small $A$} and {\small $B$} are randomly generated sparse matrices with various densities.
  In all cases the sparse matrices have a size of 10,000$\times$10,000.
  }
  \label{fig:experiments}
\end{figure}

Fig.~\ref{fig:experiments}(a) shows the time taken for unordered element insertion
done directly using the underlying storage formats (ie.,~CSC, COO, RBT, as per Section~\ref{sec:formats}),
as well as the hybrid approach which uses RBT followed by conversion to CSC.
The CSC and COO formats use oversized storage as a form of optimisation.
The RBT format is the quickest, generally by one or two orders of magnitude,
with the conversion from RBT to CSC adding negligible overhead.
The results for quasi-ordered insertion (not shown) follow a similar pattern.

Fig.~\ref{fig:experiments}(b)
shows the time taken to calculate the  expression {\small $\operatorname{trace}(A^T B)$},
with and without the aid of the automatic template-based optimisation of expression patterns,
as mentioned in Section~\ref{sec:functionality}.
Employing expression optimisation leads to considerable reduction in the time taken.
As the density increases (ie., more non-zero elements),
more time is saved via expression optimisation.

\section{Conclusion}
\label{sec:conclusion}

Motivated by a lack of easy-to-use tools for sparse matrix development, we have
proposed and implemented a sparse matrix class in C++ that internally uses a
hybrid format.  The hybrid format automatically converts between good
representations for specific functionality, allowing the user to write sparse
linear algebra without requiring to consider the underlying storage format.
Internally, the hybrid format uses the CSC (compressed sparse column), COO
(coordinate list), and RBT (red-black tree) formats.  In addition, template
meta-programming is used to optimise common expression patterns,
resulting in faster execution.
We have made our implementation available as part of the open-source Armadillo C++ library~\cite{Armadillo_JOSS_2016}.

The class has already been successfully used in open-source projects such as MLPACK,
a C++ library for machine learning and pattern recognition~\cite{MLPACK_JOSS_2018}, 
as well as the {\it ensmallen} optimisation library~\cite{ensmallen_2018}.
In both cases the sparse matrix class is used to allow various algorithms to be run on either sparse or dense datasets.
Furthermore, bindings are provided to the R environment via RcppArmadillo~\cite{RcppArmadillo_2014}.

Future avenues for exploration include integrating more specialised matrix formats~\cite{duff2017direct}
in order to automatically speed up specific operations.
An extended and revised version of this paper has been published in~\cite{SpMat_MCA_2019}.

\newpage

\bibliographystyle{splncs04}
\bibliography{refs}

\end{document}